\documentstyle[prd,aps,epsf]{revtex4b1}

\begin{document}

\twocolumn[\hsize\textwidth\columnwidth\hsize\csname
@twocolumnfalse\endcsname

\draft


\title{Light Hadron Spectrum in Quenched Lattice QCD with Staggered Quarks}

\author{Seyong Kim}
\address{Department of Physics, Sejong University, Seoul 143-747,
Korea}
\author{Shigemi Ohta}
\address{Institute for Particle and Nuclear Studies, KEK, Tsukuba,
305-0801, Japan}
 
\maketitle

\begin{abstract}\noindent
Without chiral extrapolation, we achieved a realistic nucleon to
\(\rho\)-meson mass ratio of \(m_N/m_\rho = 1.23 \pm 0.04 ({\rm
statistical}) \pm 0.02 ({\rm systematic})\) in our quenched lattice
QCD numerical calculation with staggered quarks.  The systematic error
is mostly from finite-volume effect and the
finite-spacing effect is negligible.  The flavor symmetry breaking in
the pion and \(\rho\) meson is no longer visible.  The lattice
cutoff is set at 3.63 \(\pm\) 0.06 GeV, the spatial lattice volume is
(2.59 \(\pm\) 0.05 fm)\(^3\), and bare quarks mass as low as 4.5 MeV
are used. Possible quenched chiral effects in hadron mass are
discussed.
\end{abstract}

]

\setcounter{page}{1}
\pagestyle{plain}

Reproducing the known light hadron mass spectrum is the most important
test the numerical lattice QCD is yet to pass, in spite of the steady
progress \cite{Review} since the pioneering works by Weingarten and
Hamber and Parisi \cite{Early}.  The main obstacle is the difficulty
in including light dynamical quarks, and consequently the available
full-QCD calculations still suffer from too heavy quark mass, too
coarse lattice spacing or too small lattice volume \cite{spectreview}.
On the other hand, with the quenched approximation where one neglects
dynamical quark loops, recent calculations use small enough lattice
spacing and large enough lattice volume to understand the systematic
errors arising from them \cite{spectreview,Gottlieb}.  Indeed recent
quenched calculations \cite{MILC,KS1,KS2,CPPACS} collectively have
shown that both of these errors are smaller than the statistical
noise, albeit with rather heavy quarks.  Yet these calculations left
three major problems: nucleon to \(\rho\)-meson mass ratio is too
high, pion to \(\rho\)-meson mass ratio is too high, and extrapolating
the results to more realistically light quark mass values is necessary
but difficult because of the subtle issue of the quenched chiral
effect \cite{Sharpe_etc}. Hence quenched calculations with
realistically light quark mass values on a large enough and fine
enough lattice are desirable.

Helped by the results from ref.\ \cite{MILC,KS1,KS2}, we choose a
gauge coupling of \(6/g^2 = 6.5\) and a lattice volume of \(48^3
\times 64\).  We will find later in this letter that these parameters
correspond to the lattice spacing of \(a = 0.0544(9) {\rm fm}\) or
the cutoff of \(a^{-1} = 3.63(6) {\rm GeV}\) and a spatial volume of
\((2.59(5) {\rm fm})^3\).  We use staggered quarks because it is
definitely superior to the Wilson one in controlling the quark mass and
hence in investigating the issue of quenched chiral effect: the
quark mass is well defined and protected by the remnant U(1) chiral
symmetry in the former while in the latter one encounters a difficult
problem of defining the critical hopping parameter under the
inevitable presence of exceptional gauge configurations
\cite{BDEGT}.  For the gauge part we use the single-plaquette Wilson
action because our lattice spacing is fine enough.

A combination of multi-hit Metropolis and over-relaxation algorithms
is used to generate Monte Carlo samples of quenched gauge field.
Separation between propagator sampling is 2000 such updates: 1000
Metropolis interleaved with 1000 over-relaxation.  This is proven good
enough from the auto-correlation analysis of the obtained pion
propagators.  We use the conjugate gradient (CG) method for inverting
the staggered quark Dirac matrices.  A few different sizes of corner
and even wall sources with bare mass values of $m_q a = (0.05, 0.04,
0.03, 0.02, 0.015, 0.0075)$ (Set I) and $m_q a = (0.01, 0.005, 0.0025,
0.00125)$ (Set II) are used for calculating staggered quark
propagators.  These two propagator sets are obtained from two almost
independent sets of gauge field configurations: they share only a few
gauge field configurationsin common.  This is to further reduce
correlations and should lead us to better comparison of
fitting results from one set with those from the other set.  They are all
combined with point sinks.  Set I and Set II together, the bare quark
mass varies for over a factor of 40 and provides us a good theater in
studying the chiral behaviors. We tried two different kinds of wall
definition and a few different source sizes to eliminate systematics
arising from using a single kind and size.  118 of quark propagators is
collected for Set I, and 250 for Set II.  The numerical algorithms are
basically the same as in ref.\ \cite{KS1} and technicalities associated
with our implementation on VPP-500 vector-parallel supercomputer is given
in ref.\ \cite{Kim_Ohta}.  Preliminary reports of the obtained results
were given in ref.\ \cite{Kim_Ohta_etc}.

\begin{figure}
\epsfxsize=0.8 \hsize
\epsffile{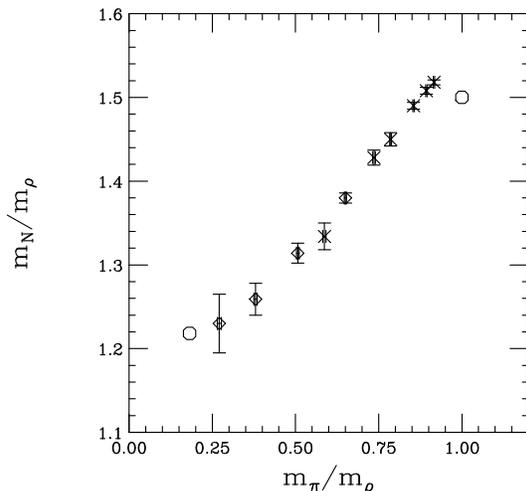}
\caption{The nucleon to $\rho$ mass ratio {\it vs}.\ pion to \(\rho\)
mass ratio at $6/g^2=6.5$ for $m_q a$ = 0.05, 0.04, 0.03, 0.02, 0.015,
0.01, 0.0075, 0.005, 0.0025, and 0.00125.  Set I is plotted with
diamonds ($\diamond$), and Set II crosses ($\times$).  Fit values with
only statistical errors are shown and no continuum or finite-volume
correction is made.  The lower circle represents the experimental
value ($(m_N/m_\rho, m_\pi/m_\rho) = (1.218, 0.182)$) and the upper
circle represents the non-relativistic limit ($ = (1.5,1.0)$).}
\label{fig:Edinburgh}
\end{figure}

Table \ref{tab:hadronmass} summarizes our estimates for the pion,
\(\rho\)-meson and nucleon mass values.
\noindent
Here \(m_\pi a \) is the mass estimate for the mass of Goldstone pion,
while \(m_{\pi_2} a\) is the estimate for non-Goldstone pion extracted
simultaneously with the estimate for its parity partner scalar
$f_0/a_0$.  Similarly \(m_\rho a\) is the estimate from the vector
meson partnered with \(b_1\) axial while \(m_{\rho_2} a\) is the
estimate extracted simultaneously with its parity partner $a_1$ 
axial.  The nucleon mass \(m_N a\) is from the even-source
results which gave better signals than the corner-source ones.
Fitting is done by minimizing the correlated $\chi^2$ calculated from
a single elimination jack-knife data set.  From these data a few
immediate conclusions follow: 1) In Fig.\ \ref{fig:Edinburgh} we show
our Edinburgh plot, $m_N/m_\rho$ vs.\ $m_\pi/m_\rho$.  For our lightest
bare quark mass of $m_q a = 0.00125$, we get $m_N/m_\rho = 1.230 \pm
0.035$ and $m_\pi/m_\rho = 0.273 \pm 0.006$.  The former is in good
agreement with the observed value, although the latter is still about
50\% larger.  2) The flavor symmetry breaking estimated by
($m_{\pi_2}a - m_\pi a$) and ($m_{\rho_2} a - m_\rho a$) is generally
small.  It decreases as we decrease the bare quark mass so much as to be
eventually hidden below statistical errors.  The symmetry is restored
well enough.

Now let us turn our attention to what we can learn about the
systematic errors.  First, the finite volume effect: from Table
\ref{tab:hadronmass} we see that $m_\rho a $ between $m_q a = 0.0025$
and $m_q a = 0.00125$ does not change within statistical error.
Assuming, rather safely, that the \(\rho\)-meson mass dependence on
the quark mass is mild in this region around $m_q a = 0.00125$ and
taking the calculated result as physical, we estimate the physical
size of our lattice cutoff \(a^{-1}\) be \(m_\rho({\rm physical})\) /
\(m_\rho(m_q a = 0.00125) = 0.7700(8)/0.212(4) \approx 3.63(6) {\rm
GeV}^{-1}\) or \(a = 0.0544(9) {\rm fm}\).  It follows that our
lattice has spatial extent of about 2.59(5) fm.  Using nucleon mass
instead of \(\rho\)-meson mass results in a consistent estimate.  It
has been argued that the lattice-QCD finite-volume effect is sensitive
to the pion Compton wavelength \(m_\pi^{-1}\) on the lattice: when it
is large compared with the lattice, the finite volume effect is
expected to fall like \(1/V\) \cite{Fukugita}, and like \(\exp(-m_\pi
L)\) otherwise \cite{Luscher}.  An extensive study made by the MILC
collaboration \cite{Gottlieb,MILC} report that the finite-volume
effect on a \((2.7 {\rm fm})^3\) lattice is smaller than 1 \%.  This
should translate into at most 1.3 \% effect for our lattice volume
of \((2.59(5) {\rm fm})^3\) when \(1/V\) dependence is assumed.  With
the \(\exp(-m_\pi L)\) dependence the effect is smaller than this
except for the case of our lightest bare quark mass of
\(m_qa=0.00125\), where the effect is expected to be slightly larger
with \(m_\pi L = 2.76(4)\).  We also compare our data with existing
data at the same \(6/g^2=6.5\) but on a smaller \(32^3 \times 64\)
lattice \cite{KS1} for \(m_q a\) = 0.01, 0.005 and 0.0025.  Here we
see a \(0.7 \pm 1.0\) \% (\(m_qa=0.01\)) to \(4.4 \pm 3.2\) \%
(\(m_qa=0.0025\)) effect in \(m_\rho a\) (heavier $m_\rho a$ on a
larger lattice), and \(4.7 \pm 1.0\) \% (\(m_q a = 0.01\)) to \(6.3
\pm 3.2\) \% (\(m_q a = 0.0025\)) effect in \(m_N a\) (lighter \(m_N
a\) on larger lattice).  These are consistent with 4.4 \% effect
expected for the \(32^3\) volume assuming the 1.3 \% effect on the
\(48^3\) volume and the \(1/V\) behavior for $32^3$ volume.  Therefore
we estimate the finite-volume effect in the current nucleon to
\(\rho\)-meson mass ratio \(m_N/m_\rho\) result is \(\sqrt{2} \times
1.3 \% \times 1.23 \simeq 2.3 \%\).

Finite-spacing effect: here we expect \(O(a^2)\) flavor breaking
effect among various definitions of staggered pions and \(\rho\)
mesons.  However as we already discussed, the breaking is hardly
visible in our data alone.  Comparison with earlier works
\cite{Gottlieb,MILC,KS1} at lower values of \(6/g^2\) (\(\le 6.2\))
reinforces this observation.  Finite lattice spacing effect in the
mass ratio like $m_N/m_\rho$ should be even smaller than that in
$m_\rho$ and $m_N$ individually.  Though there are potential ${\cal O} (a)$
effects to $m_N/m_\rho$ from the flavor symmetry breaking in
\(m_{\pi} a\) \cite{MILC}, it should be negligible as the breaking in
\(m_\pi a\) is already hardly visible.

\begin{figure}
\epsfxsize=0.8 \hsize \epsffile{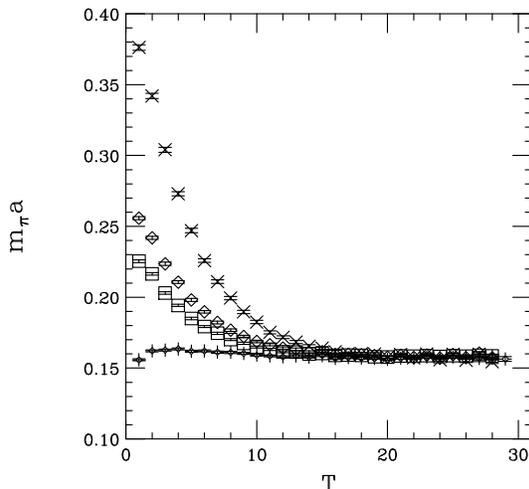}
\caption{The pion effective mass from \(12^3\) ($\times$), \(24^3\)
($\diamond$), \(32^3\) ($\Box$) and $48^3$ ($+$) wall sources at $m_q
a = 0.01$.  We see the quality of our plateau when all the four
agrees.}
\label{fig:wallsize}
\end{figure}

In addition to these systematic errors mentioned in the above, we
considered whether the size of the quark field wall source introduces
a systematic bias in choosing the best fit for hadron mass (depending
on the size of wall source, excited hadronic states can couple to the
wall differently \cite{KS1}).  For the three different wall sizes,
$12^3, 24^3$, and $32^3$, we gathered 300 hadron propagators with $m_q
a = 0.01$ and 124 hadron propagators with $m_q a = 0.00125$.  Fig.\
\ref{fig:wallsize} shows the effective mass for pion from these
different source sizes at $m_q a = 0.01$.  We see an expected tendency
that the plateau sets in later for smaller sources.  More importantly,
within the statistical error they eventually agree with each other and
eliminates ambiguity in defining a plateau.  Other effective mass
plots exhibit the same behavior and help defining plateaus.  Thus, our
choice of the best fit is less biased by a subjective choice of a
plateau.

\begin{figure}
\epsfxsize=0.8 \hsize
\epsffile{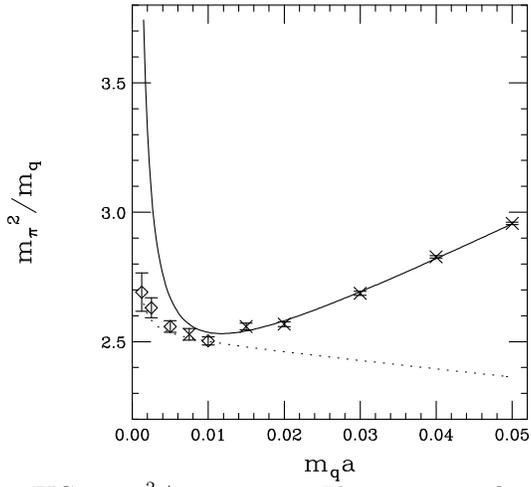}
\caption{$m_\pi^2/m_q $ {\it vs}.\  $m_q a $.  The curves are
fit to a form of $m_\pi^2/m_q = C_0/m_q + C_1 + C_2 m_q$, to  Set I
(solid) and Set II (dotted).}
\label{fig:Chiralpi}
\end{figure}

Thus we established a good enough control of the systematic errors
arising from the finite volume, finite cutoff, and choice of plateau
in the effective mass, to start discussing the quenched chiral log
problem.  In Fig.\ \ref{fig:Chiralpi}, $m_\pi^2/m_q $ vs.\ $m_q
a$ is plotted.  The fitting form, $m_\pi^2 = C_0 + C_1 m_q + C_2
m_q^2$, has been tried on Set I and Set II separately with
correlations among different $m_\pi$'s included.  This form has
been suggested by the finite volume effect on the pion mass
\cite{Mawhinney}.  At best we get $C_0 = (1.9 \pm 0.2) \times 10^{-3},
C_1 = 2.20 \pm 0.02, C_2 = 14.4 \pm 0.2 $ with confidence level (C.L.)
of $1.05 \times 10^{-5}$ (\(\chi^2/{\rm d.o.f.} = 8.6\)) for Set I,
and $C_0 = (1.7 \pm 1.1) \times 10^{-4}, C_1 = 2.51 \pm 0.04, C_2 =
-(0.30 \pm 0.33)$ with C.L. of $2.2 \times 10^{-4}$ (\(\chi^2/{\rm
d.o.f.} = 13.6\)) for Set II.  Neither fit is satisfactory.  In
particular, the fit to Set I overshoots in the quark mass region
covered by Set II.  This suggests our data in the small mass region
are much less singular than the $1/m_q $ behavior of
the finite volume effect. For Set II, a fitting form inspired by the
quenched chiral perturbation theory \cite{Sharpe_etc}, $\log
m_\pi^2/m_q = c - \delta \log m_{\pi_2}^2$ has been tried, where the
correlations among $m_\pi$ and $m_{\pi_2}$ have been fully taken into
consideration.  Fitting all the data from Set II together gives \(c =
0.815 \pm 0.037\) and \(\delta = 0.027 \pm 0.010\) with C.L. of $9.7
\times 10^{-4}$ ($\chi^2/{\rm d.o.f.} = 6.93)$.  This confidence level
or $\chi^2/{\rm d.o.f.}$ is better than that for the fitting
form considered in the above but it is still marginal.  Fitting $m_q a
= 0.005, 0.0025$ and $0.00125$ data only gives \(c = 0.691 \pm
0.052\), \(\delta = 0.057 \pm 0.013\) with C.L. of $5.8 \times
10^{-2}$ ($\chi^2/{\rm d.o.f.} = 3.59)$ while fitting to the form,
$m_\pi^2 = C_0 + C_1 m_q$, for these three quark masses gives $C_0 =
(2.29 \pm 0.98) \times 10^{-4}, C_1 = 2.47 \pm 0.02$ with C.L. of $1.6
\times 10^{-4} (\chi^2/{\rm d.o.f.} = 14.2)$.  One can try to add one
more parameter to the above quenched chiral log fit, in order to fit
all the data from Set II. Indeed, such a fit gives an improved
C.L. ($7.0 \times 10^{-3}$) but similar modification (from $C_0 + C_1
m_q$ to $C_0 + C_1 m_q + C_2 m_q^2$) does not improve C.L. of the
finite volume fitting form. Thus we think that the quenched chiral
fitting form describes our data better and the quenched chiral
logarithm behavior discussed in the $m_\pi^2/m_q $ in
\cite{KS2,CPPACS} is not a finite lattice volume artifact.

For nucleon and $\rho$-meson mass values, various fitting forms are
suggested and tried in ref.\ \cite{MILC}, where origins of each terms
in these fitting forms were discussed.  Although these forms should be
considered only after the continuum limit is taken, we try these
fitting forms on our data since finite lattice spacing effect on our
data is small as we discussed in the above.  $m_q^{1/2}$ and $m_q \log
m_q$ terms are from quenched chiral perturbation theory consideration,
and $m_q$ (from tree level), $m_q^{3/2}, m_q^2, m_q^2 \log m_q$ (from
one loop correction) terms are present both in quenched and ordinary
chiral perturbation theory.  Following them, we studied chiral
extrapolation in $m_\rho a$ and $m_N a$ using hadron masses from Set I
and compared directly with those masses from Set II.  Since Set II has
four data points, we also tried fits to hadron masses from Set II when
the number of fitting parameters is less than four.  Correlation among
hadron masses is included in the fitting.  For Set I, among the twelve
fitting forms suggested in ref. \cite{MILC}, we can definitely rule
out $a + b m_q^{1/2}$ and $a + b m_q$ because the confidence levels
are so poor.  Fitting to $a + b m_q^{1/2} + c m_q + d m_q^{3/2}$ and
$a + b m_q^{1/2} + c m_q + d m_q^2$ return either $d$ with error more
or less equal to $d$ suggesting that the d term is not necessary, or a
negative $d$.  On the other hand, the fit to $a + b m_q^{1/2} + c m_q$
gives a positive value for $b$, which is inconsistent with quenched
chiral perturbation theory \cite{qBaryon}.  Similarly, fits to $a + b
m_q + c m_q^{3/2} + d m_q^2$, $a + b m_q^{1/2} + c m_q + d m_q \log
m_q$, and $a + b m_q + c m_q^2 + d m_q^2 \log m_q$ give either $d$
with error as large as $d$ or larger than $d$ implying that fitting
forms of $a + b m_q + c m_q^{3/2}$, $a + b m_q^{1/2} + c m_q$, and $a
+ b m_q + c m_q^2$ are preferred, or a negative value for $d$.  All
the fits to $a + b m_q + c m_q^{3/2}$ and $a + b m_q^{1/2} + c m_q$
returns a negative $c$.  The fit to $a + b m_q + c m_q \log m_q$ gives
a negative $c$.  The confidence level of the fit to $a + b m_q + c
m_q^2 \log m_q$ is $7.7 \times 10^{-4}$ for $m_N a$ and $1.3 \times
10^{-2}$ for $m_\rho a$.  None of these extrapolations agrees well
with results from Set II.  In contrast, all the fitting forms except
for the $a+b m_q$ to Set II give high C.L., telling us that the
associated statistical errors of Set II are too large.  Following
ref. \cite{MILC}, we also tried fitting $m_N + \lambda_N m_{\pi_2}$
and $m_\rho + \lambda_\rho m_{\pi_2} $ to $a + b m_q + c m_q^{3/2} + d
m_q^2$ or $a + b m_q + c m_q^{3/2}$ for various $\lambda$'s but none
of them improves confidence level significantly.

\begin{figure}
\epsfxsize=0.8 \hsize
\epsffile{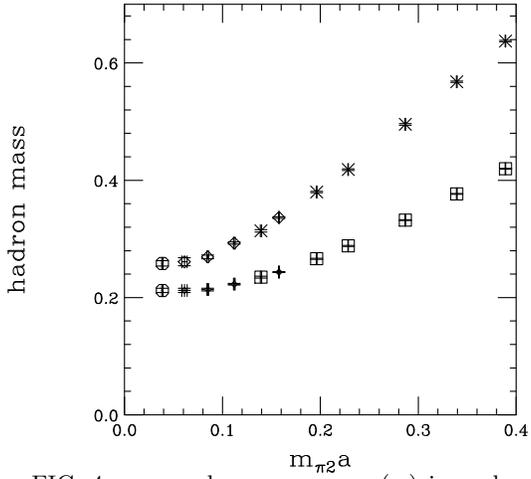}
\caption{$m_N $ and $m_\rho$ {\it vs}.  $m_{\pi_2} $.
$(\times)$ is nucleon mass from Set I and $(\Diamond)$ is nucleon mass
from Set II.  $(\Box)$ is $\rho$ mass from Set I and $(+)$ is $\rho$
mass from Set II.  The leftmost circles show
expected $m_N a$ and $m_\rho a$ for physical quark mass which is
deduced by the nucleon mass and the pion mass from experiment and by
the current lattice spacing, $a$.}
\label{fig:Chiraln2}
\end{figure}

In Fig.\ \ref{fig:Chiraln2}, our $m_N a$ and $m_\rho a$ are plotted
against $m_{\pi_2}$.  The leftmost circles show expected $m_N a$ and
$m_\rho a$ for physical quark mass which is deduced by the nucleon
mass and the pion mass from experiment and by the lattice spacing,
$a$, obtained in the above.  Quenched chiral perturbation theory
suggests a linear dependence on $m_{\pi_2}$ with a negative
coefficient.  A fit gives a positive coefficient with poor confidence
level, ${\cal O} (10^{-4})$ although the fit looks good to eyes. Our
data in Fig.\ \ref{fig:Chiraln2} clearly show that the coefficient of
linear term is small, in agreement with small $\delta $ (0.02 $\sim$
0.06) from our $m_\pi^2/m_q$ data.  The smallness of the coefficient
for the linear $m_{\pi_2}$ term and the disagreement between the
extrapolated results from Set I and the actual Set II data, lead us to
conclude that one needs very high statistics to trust chiral
extrapolation of hadron mass data when lattice calculation is
performed with heavy quark masses in order to obtain a result similar
to MILC collaboration \cite{MILC}. It is interesting that for $m_q a <
0.005$, the obtained values of pion and
\(\rho\)-meson mass would allow $\rho \rightarrow \pi\pi$
decay. Influence of such decay mode on $\rho$-meson mass needs further
investigation in the context of quenched approximation.

In conclusion our numerical calculation of quenched QCD with
\((2.59(5) {\rm fm})^3\) spatial volume and 3.6 GeV cutoff yields a
realistic mass ratio of $m_N/m_\rho = 1.230 \pm 0.035 \pm 0.023$ at
the bare quark mass of $m_q a$ = 0.00125 $\simeq$ 4.5 MeV, where the
first error is statistical and the second is finite-volume. Finite
lattice spacing effect is negligible.  This result is obtained without
chiral extrapolation.  Flatness of $m_N a$ and $m_\rho a$ in the
region $m_q a \le 0.005$ implies that there will be little variation
in lattice simulation of $m_N a$ and $m_\rho a$ from our lightest
quark mass to physical quark mass.  Concerned with quenched chiral
perturbation behavior, we tried chiral extrapolations for $m_\pi,
m_\rho$ and $m_N$ from heavier quark mass ($m_q a = 0.0075 \sim 0.05$)
(these fitting forms make sense only in the continuum limit. However,
since our lattice spacing is quite small, we tried various fitting
forms assuming that there are little modification on the fitting forms
due to the finite lattice spacing effect) and compared extrapolated
values with our simulated result with lighter quark mass ($m_q a =
0.00125 \sim 0.01$).  For ${m_\pi}^2/m_q $, although a finite lattice
volume argument suggests a singular $1/m_q$ behavior, comparison of a
fit using ${m_\pi}^2$ from heavier quark mass with ${m_\pi}^2$ from
lighter quark mass shows that $m_\pi^2/m_q $ from lighter quark mass
calculation is less singular than $1/m_q$.  In nucleon and
\(\rho\)-meson, unlike in pion, it is hard to distinguish by chiral
extrapolation the linear dependence on
\(m_{\pi_2}\), a term expected by quenched chiral perturbation theory,
because 1) the statistical fluctuation associated with $m_N a$ and
$m_\rho a$ is larger than that with $m_{\pi}$, and 2) the quenched
chiral perturbation parameter, $\delta$, appears to be smaller than
that in ref.\ \cite{MILC} probably due to our larger $6/g^2$.  This
$6/g^2$ dependence may be understandable since $\delta$ is related to
$m_0^2/(4\pi f_\pi^2)$ \cite{Sharpe_etc} and the asymptotic scaling of
$m_0$ can be different from that of $f_\pi$.

We would like to thank S.~Sharpe and D.K.~Sinclair for helpful
discussions.  We also thank the Institute for Physical and Chemical
Research (RIKEN) for the use of its Fujitsu VPP-500/30 vector-parallel
super computing system.  S.K.\ would like to thank Prof.~E.~Goto for the
hospitality during his visits to RIKEN.  S.O.\ would like to thank
the RIKEN BNL Research Center where he stayed from time to time during the
course of this work.

\begin{table}[b]
\caption{Bare quark mass \protect\(m_q\protect\) and hadron mass, all
in lattice units.  Under the staggered quark formalism the
Nambu-Goldstone (NG) pion \protect\(\pi\protect\) and non-NG pion
\(\pi_2\protect\) split at \protect\(O(a^2)\protect\) because of the
flavor symmetry breaking, and so do the \(\rho\)-mesons
\protect\(\rho\protect\) and \protect\(\rho_2\protect\), but the
effects are now so small and hardly visible.  The nucleon is from the
even sources which give better signal than the corner ones.}
\label{tab:hadronmass}
\begin{tabular}{lccccc}
$m_q a$   &$m_\pi a$  &$m_{\pi_2} a$&$m_\rho a$ &$m_{\rho_2} a $&$m_N a$ \\
\hline
0.05    &0.3845(4)&0.3890(4) &0.4196(6) &0.4198(5)  &0.637(1) \\
0.04    &0.3363(4)&0.3394(4) &0.3767(6) &0.3770(6)  &0.568(1) \\
0.03    &0.2839(4)&0.2868(4) &0.3322(7) &0.3316(7)  &0.495(1) \\
0.02    &0.2266(5)&0.2284(5) &0.2882(9) &0.2878(9)  &0.418(2) \\
0.015   &0.1959(6)&0.1962(5) &0.2661(10)&0.2664(11) &0.380(2) \\
0.01    &0.1582(5)&0.1577(5) &0.2434(8) &0.2417(9)  &0.336(1) \\
0.0075  &0.1377(6)&0.1394(7) &0.2347(16)&0.2367(15)&0.313(3) \\
0.005   &0.1131(6)&0.1121(8) &0.2229(13)&0.2225(13) &0.293(2) \\
0.0025  &0.0811(7)&0.0850(10)&0.2137(22)&0.2122(21) &0.269(3) \\
0.00125 &0.0576(8)&0.0612(29)&0.2122(37)&0.2117(33) &0.261(6)
\end{tabular}
\end{table}

\end{document}